\documentclass[aps,showpacs,twocolumn]{revtex4}

\usepackage[dvips]{graphics,graphicx}
\usepackage{amsmath}
\usepackage{amssymb}                 
\usepackage{amscd}                   
\usepackage{wasysym}
\usepackage[ansinew]{inputenc}
\usepackage[T1]{fontenc}
\usepackage{ae,aecompl}

\newcommand{\ri}{i}
\newcommand{\re}{{ \rm e }}
\newcommand{\rd}{d}
\newcommand{\sk}{{\rm sk}}
\newcommand{\rC}{{C}}
\newcommand{\rA}{{A}}

\newcommand{\be}{\begin{equation}}
\newcommand{\ee}{\end{equation}}

\renewcommand{\vec}[1]{\mathbf{#1}}

\usepackage{color}
\definecolor{blau}{rgb}{0,0,1}
\definecolor{gruen}{rgb}{0,1,0}
\definecolor{rot}{rgb}{1,0,0}
\definecolor{magenta}{rgb}{1,0,1}

\begin{document}
\bibliographystyle{apsrev}
\title{Barrier transmission for the one--dimensional nonlinear Schr\"odinger equation: resonances and transmission profiles}
\author{K. Rapedius and H. J. Korsch}
\email{korsch@physik.uni-kl.de}
\affiliation{Technische Universit{\"a}t Kaiserslautern, FB Physik,
                            D-67653 Kaiserslautern, Germany}
\date{\today }

\begin{abstract}
The stationary nonlinear Schr\"odinger equation (or Gross-Pitaevskii equation)
for one-dimensional potential scattering is studied. The nonlinear
transmission function shows a distorted profile, which differs from the
Lorentzian one found in the linear case. This nonlinear profile
function is analyzed and related to Siegert type complex resonances.
It is shown, that the characteristic nonlinear profile function can be
conveniently described in terms of skeleton functions depending
on a few instructive parameters. These skeleton functions also determine the decay behavior of the underlying resonance state. Furthermore we extend the Siegert method
for calculating resonances, which provides a
convenient recipe for calculating nonlinear resonances. Applications
to a double Gaussian barrier and a square well potential illustrate
our analysis.

\noindent

\end{abstract}
\maketitle

\section{Introduction}
Transport properties of Bose-Einstein condensates (BECs) are of
considerable current interest, both experimentally and theoretically. Especially atom--chip experiments are well--suited to study the influence of interatomic interaction on transport properties of BECs in wavegiudes since different waveguide geometries can easily be realized \cite{Folm00,Hans01,Ott01,Ande02}. An alternative method was implemented in a recent experiment \cite{Guer06} where a BEC was created in an optomagnetic trap and outcoupled into an optical waveguide.

A convenient theoretical approach is based on the Gross-Pitaevskii equation (GPE)
or nonlinear Schr\"odinger equation (NLSE)
\be
  \ri \hbar \frac{\partial \psi(\vec{x},t)}{\partial t}  =
  \left( -\frac{\hbar^2}{2 m}  \nabla^2 + V(\vec{x})
  + g | \psi(\vec{x}, t) |^2 \right)  \psi(\vec{x},t) \, ,
   \label{GPE}
\ee
which describes the dynamics in a mean-field approximation at low
temperatures \cite{Pita03,Park98,Dalf99,Legg01}. Another important application of the NLSE is the propagation of electromagnetic waves in nonlinear media (see, e.g., \cite{Dodd82}, ch. 8).
The ansatz $\psi({\bf x},t)=\exp(-\ri \mu t/\hbar)\,\psi(\bf x)$ reduces
(\ref{GPE}) to the
corresponding time-independent NLSE
\be
  \left( -\frac{\hbar^2}{2 m}  \nabla^2 + V(\vec{x})
  + g | \psi(\vec{x}) |^2 \right)  \psi(\vec{x})
  =\mu \psi(\vec{x})
   \label{NLSE_stat}
\ee
with the chemical potential $\mu$.

\begin{figure}[htb]
\centering
\includegraphics[width=7cm,  angle=0]{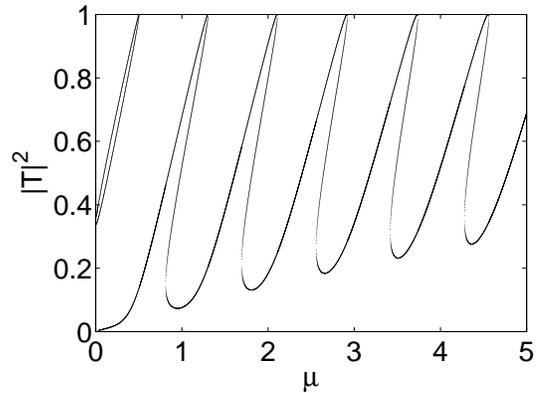}
\caption{\label{fig-Tq_bsp} {Transmission coefficient obtained from solving the stationary NLSE (\ref{1}) for a square well potential with  $a=20$, $V_0=-50$ for $g=+1$ (see Sec.~\ref{subsec_sq} and \cite{06nl_transport}).
}}
\end{figure}

Various interesting phenomena have been reported originating from the
nonlinearity of Eq.~(\ref{GPE}), as for instance a bistability of the
barrier transmission probability \cite{Paul05,Paul05b,04nls_delta,06nl_transport}.
A paradigmatic model in this context is the transmission through
a one-dimensional rectangular barrier or across a square well potential, one
of the rare cases were the one--dimensional NLSE
\be
    \frac{\hbar^2}{2m} \psi''+(\mu-V)\psi-g|\psi|^2\psi=0
    \label{1}
\ee
can be solved analytically \cite{Carr00a,Carr00b,06nl_transport,05dcomb}. As an example, Fig.~\ref{fig-Tq_bsp}
shows the nonlinear transmission coefficient $|T|^2$ as a function of the chemical potential for the square well potential considered in \cite{06nl_transport}, which is discussed in more detail in section \ref{sec_app}.

One observes a clear structured behavior: the well-known Lorentz profiles
determined by the complex-valued resonances, well understood
for linear transmission, are distorted. The curves bend to the right
(to the left for attractive nonlinearity $g<0$) and are multivalued in
certain regions.
A time--dependent numerical analysis shows that the lowest branch of the transmission coefficient is populated if an initially empty waveguide is slowly filled with condensate with a fixed chemical potential $\mu$ \cite{06nl_transport,Paul05} whereas the highest branch can be populated if the chemical potential is adiabatically increased during the propagation process which is equivalent to applying an additional weak time--dependent potential \cite{Paul05}. These and related aspects are discussed in detail in \cite{Paul07}.

It is the purpose of the present paper to determine the functional form of the transmission profile surrounding resonance peaks characterized by $|T|^2=1$ for a general (symmetric) potential
by relating them to Siegert type complex resonances. In particular we show that the lineshape of the resonance peaks is determined by the decay behavior of the underlying metastable resonance state.

The paper is organized as follows. In Sec.~\ref{sec-nl_Lorentz}, we develop a formula for the nonlinear Lorentz profiles which describes the transmission coefficient in the vicinity of a resonance in terms of skeleton functions which also determine the decay behavior of the resonance. In Sec.~\ref{sec_skel}, we present a convenient recipe for calculating nonlinear resonances and skeleton curves which we call the Siegert method. In Sec.~\ref{sec_app}, we use this method to demonstrate the validity of the nonlinear Lorentz profile from Sec.~\ref{sec-nl_Lorentz} for two example potentials. In Appendix \ref{app_Continuation} we discuss the continuation of solutions of the NLSE to complex chemical potentials and in Appendix \ref{app_SiegertRel} we derive a useful formula for the decay coefficient of a symmetric finite range potential which we call the Siegert relation.

\section{Nonlinear Lorentz profile}
\label{sec-nl_Lorentz}
\subsection{The transmission problem}
\label{subsec_Transmission}
\begin{figure}[htb]
\centering
\includegraphics[width=9cm, angle=0]{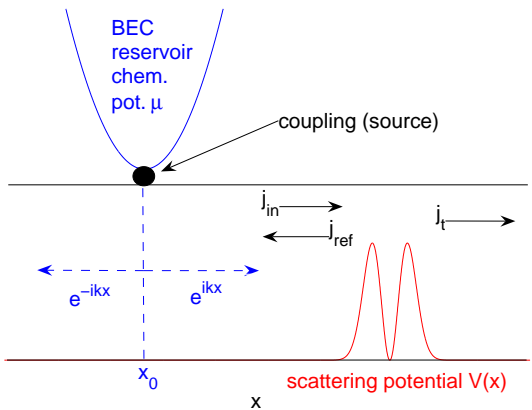}
\caption{\label{fig-source} {(Color online) At $x_0$ condensate with chemical potential $\mu$ is injected into the waveguide from a reservoir. The reservoir emits a plane matter wave in both directions into the guide so that an incoming beam with current $j_i$ is partially reflected (current $j_{ref}$ and partially transmitted (current $j_t$) at the barrier potential $V(x)$.
}}
\end{figure}
In the case of the NLSE the superposition principle is not valid. Therefore the definition of a transmission coefficient is nontrivial. Here we review and slightly extend an approach based on the time--dependent NLSE (see \cite{Paul05,Paul07,06nl_transport}). Following Paul et al.~\cite{Paul07} we consider an experimental setup where matter waves from a large reservoir of condensed atoms at chemical potential $\mu$ are injected into a one-dimensional waveguide in which the condensate can propagate (see Figure \ref{fig-source}).
In a time--dependent approach the system is described by the NLSE
\begin{eqnarray}
   \ri \hbar \dot{\psi}(x,t)=&-&\frac{\hbar^2}{2m}\psi''(x,t)+V(x)\psi(x,t) \\ \nonumber
   &+&g|\psi(x,t)|^2\psi(x,t)+f(t)\re^{-\ri \mu t/\hbar}\delta(x-x_0)
   \label{GPE_t}
\end{eqnarray}
where the source term $f(t)\re^{-\ri \mu t/\hbar}\delta(x-x_0)$ located at $x=x_0$ emits monochromatic matter waves at chemical potential $\mu$ and thus simulates the coupling to a reservoir.
The barrier potential $V(x)$ is assumed to be zero for $x \le x_0$.

In the following we assume a constant source strength $f(t)=f_0$ and look for stationary solutions $\psi(x,t)=\psi(x)\exp(-\ri \mu t/\hbar)$ of Eq.~(\ref{GPE_t}) arriving at
\begin{eqnarray}
   \mu \psi(x)= &-& \frac{\hbar^2}{2m}\psi''(x)+V(x)\psi(x) \\ \nonumber
   &+&g|\psi(x,t)|^2\psi(x,t)+f_0 \delta(x-x_0) \, .
\end{eqnarray}
Application of the integral operator
\be
                    \begin{array}{cc}
                     {\rm lim}  \\
                    \epsilon \rightarrow 0\\
                     \end{array}   \int_{x_0-\epsilon}^{x_0+\epsilon} \rd x
\ee
leads to
\be
   -\frac{\hbar^2}{2m}(\psi_+'-\psi_-')\Big|_{x_0}+f_0=0
   \label{rel1}
\ee
where we have introduced the notation
\be
  \psi(x)= \left\{
                    \begin{array}{cl}
                     \psi_-(x)  &   x \le x_0 \\
                     \psi_+(x)  &   x> x_0\\
                     \end{array}
              \right. \, .
\ee
Since we have $V(x)=0, x \le x_0$ and there is no incoming current from $x=-\infty$ the solution in the region $x \le x_0$ is given by the plane wave $\psi_-(x)=\psi_-^0 \exp(-\ri k_- x)$ with $k_-=\sqrt{2m(\mu-g|\psi_-^0|^2)}/\hbar$ and $\psi_-'(x)=-\ri k_- \psi_-(x)$.
Inserting this into Eq.~(\ref{rel1}) leads to
\be
   f_0=\frac{\hbar^2}{2m}(\psi_+'+\ri k_- \psi_-)\Big|_{x_0}=0 \\
\ee
or, taking into account the continuity of the wavefunction $(\psi_+-\psi_-)\big|_{x_0}=0$ at $x=x_0$, to
\be
   f_0=\frac{\hbar^2}{2m}(\psi_+'+\ri k_+ \psi_+)\Big|_{x_0}=0
   \label{rel3}
\ee
with $k_+=\sqrt{2m(\mu-g|\psi_+(x_0)|^2)}/\hbar$.
Eq.~(\ref{rel3}) relates the wavefunction $\psi_+$ in the region $x \ge x_0$ to the source strength $f_0$. In order to relate the source strength with the incoming condensate current we consider the
special case where $V(x)=0$ everywhere, i.e. without a barrier. Then the wave function in the region $x \ge x_0$ is given by a plane wave $ \psi_+(x)=A \exp(\ri k_A(x-x0))$ with $k_A=\sqrt{2m(\mu-g|A|^2)}/\hbar$ and $\psi_+'(x)=-\ri k_A \psi_+(x)$.  From the continuity of the wave function $\psi_-(x_0)=\psi_+(x_0)=A$ we get $k_-=k_+=k_A$ and together with Eq.~(\ref{rel3}) we obtain
\be
   f_0=\ri \frac{\hbar^2}{m}k_A A \, .
   \label{f0A}
\ee
For a given source strength $f_0$, Eq.~(\ref{f0A}) can have up to two different solutions for $A$. In the following we only consider the solution corresponding to the limit of weak interaction.
The incoming current emitted by the source is given by $j_{in}=\frac{\hbar}{m}k_\rA|A|^2=|f_0||A|/\hbar$.
Inserting Eq.~(\ref{f0A}) into Eq.~(\ref{rel3}) we obtain the relation
\be
   2 \ri k_\rA A=(\psi_+'+\ri k_+ \psi_+)\Big|_{x_0}=0 \label{2ikA}
\ee
connecting the condensate wavefunction with the incoming current.
We define the transmission coefficient as
\be
   |T|^2=\frac{j_t}{j_ {in}}
\ee
where the current $j_t$ transmitted through the barrier is obtained by evaluating the current operator
\be
   j_t=-\frac{\ri \hbar}{2m}(\psi_+^*\psi_+'-\psi_+\psi_+'^*)
\ee
anywhere in the region $x \ge x_0$, and $j_{in}$ is the current in absence of the barrier. In the noninteracting limit $g \rightarrow 0$ this definition coincides with the usual definition of the transmission coefficient known from the linear Schr\"odinger theory. It has the advantage of being applicable in the time--dependent case.
An alternative way to extend the concept of transmission to the interacting case is discussed in \cite{Paul07}.
\subsection{Resonance lineshape}
In the following we will derive a formula for the transmission coefficient $|T|^2$ of a symmetric potential well $V(x)=V(-x)$ with the finite range $a$ (i.e. $V(x)=0$ if $|x|>a$) in the vicinity of a resonance in dependence of the chemical potential $\mu$ of the incoming condensate current. Our approach is based upon a generalization of Siegert's derivation of the dispersion formula for nuclear reactions and we closely follow the arguments in \cite{Sieg39}.
An alternative ansatz makes use of the Feshbach formalism \cite{Schl06H,Paul07}.

We consider the situation where the condensate source is located at $x_0=-a$. The solution in the downstream region $x>a$ is then given by a plane wave $\psi(x)=C \exp(\ri k_\rC x)$ with $k_\rC=\sqrt{2m(\mu-g|C|^2)}/\hbar$ so that the transmitted current is given by $j_t=\frac{\hbar}{m}k_\rC|C|^2$. At $x=-a$ the wave function must satisfy Eq.~(\ref{2ikA}) with $x_0=-a$.

Thus the scattering wavefunction $\psi$ in the interval $[-a,a]$ is a solution of Eq.~(\ref{1})
with boundary conditions
\begin{eqnarray}
&&\psi(a)=C \re^{\ri k_\rC a},\quad \: \psi'(a)=\ri k_\rC \psi(a)  \label{2a}\\
&& 2 \ri k_\rA A=\psi'(-a)+\ri k \psi(-a)=0
\label{2c}
\end{eqnarray}
where $k=\sqrt{2m(\mu-g|\psi(-a)|^2)}/\hbar$.
The transmission coefficient given by $|T|^2=j_t/j_{in}=(k_\rC/k_\rA) |C/A|^2$ depends on the chemical potential $\mu$ and, due to the nonlinear term in (\ref{1}), also on the magnitude of the wavefunction. If the incoming amplitude $A$ is kept fixed, this dependence can be conveniently described by the magnitude $|C|^2$ of the outgoing amplitude.

Now we consider $C/A$ as a function of $\mu$. From (\ref{2a}) - (\ref{2c}) we obtain
\be
   \frac{C}{A}=\frac{2 \ri k_\rA \psi(a)\exp(-\ri (k+k_\rC) a)}{\ri k \psi(-a)+\psi'(-a)} \, .
   \label{3}
\ee
Singularities of (\ref{3}) occur for certain complex chemical potentials $W_\sk$ where the denominator vanishes. These values of the chemical potential are defined by the eigenvalue problem
\be
   \frac{\hbar^2}{2m} \psi_\sk''+(W_\sk-V)\psi_\sk-g|\psi_\sk|^2\psi_\sk=0
    \label{4}
\ee
in $-a \le x \le a$ with the boundary conditions
\begin{eqnarray}
\psi_\sk(a)=C \re^{\ri k_\sk a},\quad \quad \: \, & \psi_\sk'(a)=\ri k_\sk \psi_\sk(a)\quad \quad & \label{5a}\\
\psi_\sk(-a)=C \re^{\ri (k_\sk a+ \ri \delta)}, \, & \psi_\sk'(-a)=-\ri k_\sk \psi_\sk(-a) \quad&
\label{5}
\end{eqnarray}
where $k_\sk=\sqrt{2m(W_\sk-g|C|^2)}/\hbar$ and $\delta$ is some real valued phase. Because of the nonlinear term in (\ref{4}), the complex energy $W_\sk=\mu_\sk-\ri \Gamma_\sk/2$ with real $\mu_\sk$ and $\Gamma_\sk$ depends explicitly on $|C|^2$. The problem concerning the continuation of the solution to the domain of complex chemical potentials is discussed in Appendix \ref{app_Continuation}.
Motivated by the analogy to a driven nonlinear oscillator
we call the functions $\mu_\sk(|C|^2)$ and $\Gamma_\sk(|C|^2)$ {\it skeleton curves} and $\psi_\sk$ the {\it skeleton wavefunction}.

From Eq.~(\ref{4}) and its complex conjugate as well as the boundary conditions (\ref{5a}), (\ref{5}) we derive the useful formula
\be
   \frac{\hbar^2}{2m}|\psi_\sk(a)|^2=\frac{\Gamma_\sk/2}{k_\sk+k_\sk^*}\int_{-a}^a|\psi_\sk|^2 \rd x \, .
   \label{15}
\ee

In order to obtain $C/A$ in the vicinity of the singularity $W_\sk$ we multiply (\ref{1}) by $\psi_\sk$ and (\ref{4}) by $\psi$ and subtract these equations. By integrating the resulting equation
\begin{eqnarray}
    \frac{\hbar^2}{2m} \big( \psi_\sk''  \psi&-&\psi''\psi_\sk \big) + (W_\sk-\mu)\psi_\sk\psi \nonumber \\
     &+& g\psi_\sk \psi (|\psi|^2-|\psi_\sk|^2)=0
\label{7}
\end{eqnarray}
from $x=-a$ to $x=+a$ and using the boundary conditions we arrive at
\begin{eqnarray}
   &&\frac{\hbar^2}{2m} \Big[ \ri \psi_\sk(a) \psi(a)(k_\sk-k_\rC) \quad \quad \nonumber \\
   && \quad \quad +\psi_\sk(-a) (\ri k_\sk \psi(-a)+\psi'(-a)) \Big] \nonumber \\
   && \quad \quad + g \int_{-a}^a \psi_\sk \psi (|\psi|^2-|\psi_\sk|^2) \rd x  \nonumber \\
   && \quad \quad + (W_\sk-\mu) \int_{-a}^a \psi_\sk\psi \rd x=0 \, .
   \label{9}
\end{eqnarray}
Thus we can write the denominator of $C/A$ in (\ref{3}) as
\begin{eqnarray}
\ri k \psi(-a)&+&\psi'(-a)= \nonumber \\
 &-&\frac{W_\sk-\mu}{(\hbar^2/2m)\psi_\sk(-a)} \int_{-a}^{a} \psi_\sk\psi \rd x \nonumber \\
 &-&\frac{2mg}{\hbar^2\psi_\sk(-a)}\int_{-a}^a \psi_\sk \psi (|\psi|^2-|\psi_\sk|^2) \rd x \nonumber \\
 &-&\ri (k_\sk-k_\rC) \frac{\psi_\sk(a) \psi(a)}{\psi_\sk(-a)} \nonumber \\
 &-& \ri(k_\sk-k)\psi(-a)  \, .
 \label{10}
\end{eqnarray}
Using
\begin{eqnarray}
   k_\sk-k_\rC &=& \frac{2m}{\hbar^2}\frac{W_\sk-\mu}{k_\sk+k_\rC} \: \: \\
   k_\sk-k &=& \frac{2m}{\hbar^2}\frac{W_\sk-\mu-g(|C|^2-|\psi(-a)|^2)}{k_\sk+k}
   \label{11}
\end{eqnarray}
we obtain
\begin{eqnarray}
&&\ri k \psi(-a)+\psi'(-a)= \nonumber \\
 &&\quad -\frac{W_\sk-\mu}{(\hbar^2/2m)\psi_\sk(-a)} \int_{-a}^{a} \psi_\sk\psi \rd x  \\
 &&\quad -\frac{2mg}{\hbar^2\psi_\sk(-a)}\int_{-a}^a \psi_\sk \psi (|\psi|^2-|\psi_\sk|^2) \rd x \nonumber \\
&&\quad  -\ri \frac{W_\sk-\mu}{(\hbar^2/2m) \psi_\sk(-a)}\frac{\psi_\sk(a)\psi(a)}{k_\sk+k_\rC} \nonumber \\
 && \quad -\ri \frac{W_\sk-\mu-g(|C|^2-|\psi(-a)|^2)}{(\hbar^2/2m) \psi_\sk(-a)}\frac{\psi_\sk(-a)\psi(-a)}{k_\sk+k} \,. \nonumber
 \label{12}
\end{eqnarray}
Assuming that the eigenvalue $W_\sk$ is not degenerate, we have in the limit $\mu \rightarrow W_\sk$: $\psi \rightarrow \psi_\sk$, $|\psi(-a)|^2 \rightarrow |C|^2$ and $k,k_\rC \rightarrow k_\sk$ so that
\begin{eqnarray}
&&\ri k \psi(-a)+\psi'(-a) \rightarrow -\frac{W_\sk-\mu}{(\hbar^2/2m) \psi_\sk(-a)}  \nonumber \\
 &&\quad \quad \quad \times \Big[ \int_{-a}^{a} \psi_\sk^2 \rd x + \ri \frac{\psi_\sk^2(a)+\psi_\sk^2(-a)}{2k_\sk} \Big] \, .
  \label{13}
\end{eqnarray}
For the numerator of $C/A$ we get $2 \ri k_\rA \psi(a) \exp(-\ri (k+k_\rC)a) \rightarrow 2 \ri k_\rA \psi_\sk(a) \exp(-2 \ri k_\sk a)$.
Thus $C/A$ becomes
\begin{eqnarray}
\frac{C}{A}=-\frac{\hbar^2/2m}{W_\sk-\mu}\frac{2 \ri k_\rA \re^{-2\ri k_\sk a}\psi_\sk(a)\psi_\sk(-a)}{\int_{-a}^{a} \psi_\sk^2 \rd x + \ri \frac{\psi_\sk^2(a)+\psi_\sk^2(-a)}{2k_\sk}} \, .
\label{14}
\end{eqnarray}

For sufficiently small values of $\Gamma_\sk$, we can multiply $\psi_\sk$ by a suitable constant $\zeta$ of magnitude $1$ which makes $\zeta \psi_\sk$ real (up to terms of order $\Gamma_\sk$) in the regions of slowly varying phase which give the main contribution to the integral $\int_{-a}^a |\psi_\sk|^2 \rd x$. In this limit we can thus use the approximation
\be
   \int_{-a}^a \zeta^2 \psi_\sk^2 \rd x \approx \int_{-a}^a |\psi_\sk|^2 \rd x \, .
   \label{15b}
\ee
We furthermore define the phase factor $\delta_\sk$
by $\zeta \psi_\sk(a)=|\psi_\sk(a)|\exp(\ri k_\sk a+\ri \delta_\sk/2)$.
Using these definitions and $\psi_\sk(-a)=\psi_\sk(+a) \exp(\ri \delta)$ Eq.~(\ref{14}) can be written as
\begin{eqnarray}
\frac{C}{A} &=&- \frac{\hbar^2/2m}{W_\sk-\mu}  \\
&\times& \frac{2 \ri k_\rA \exp(\ri \delta_\sk+\ri \delta)|\psi_\sk(a)|^2}{\int_{-a}^{a} |\psi_\sk|^2 \rd x + \ri |\psi_\sk(a)|^2 \frac{\exp(2\ri k_\sk a + \ri \delta_\sk)(1+\exp(\ri \delta))}{2 k_\sk}} \nonumber
\label{16}
\end{eqnarray}
or, using (\ref{15}),
\begin{eqnarray}
\frac{C}{A} &=&- \re^{\ri \delta_\sk+\ri \delta}\frac{\Gamma_\sk/2}{W_\sk-\mu}  \label{17}\\
&\times& \frac{2 \ri k_\rA }{k_\sk+k_\sk^* + \ri \Gamma_\sk \frac{2m \exp(2\ri k_\sk a + \ri \delta_\sk)(1+\exp(\ri \delta))}{\hbar^2 k_\sk}} \nonumber \, .
\end{eqnarray}

As $\Gamma_\sk$ tends towards zero, $k_\sk$ becomes real and the last term in the denominator of (\ref{17}) becomes negligible, so that we have in this limit
\begin{eqnarray}
\frac{C}{A} = \re^{\ri \delta_\sk+\ri \delta}\frac{k_\rA}{k_\sk}\frac{\ri \Gamma_\sk/2}{\mu-\mu_\sk+\ri \Gamma_\sk/2} \, .
\label{18}
\end{eqnarray}
Thus the transmission coefficient in the vicinity of a resonance is given by $|T|^2=(k_\rC/k_\rA) |C/A|^2$
\be
  |T|^2= \frac{k_\rA}{k_\sk}\frac{\Gamma_\sk^2/4}{(\mu-\mu_\sk)^2+\Gamma_\sk^2/4} \,
  \label{Lor_nl}
\ee
with $k_\sk \approx k_\rC$.
This result formally resembles the Lorentz or Breit-Wigner form that occurs in the respective linear theory. However the chemical potential $\mu_\sk$ and the width $\Gamma_\sk$ depend implicitly on $|T|^2$. This dependence disappears in the linear limit $g \rightarrow 0$  and we recover the usual Lorentz profile.
Eq.~(\ref{Lor_nl}) can be inverted to the form
\be
   \mu_\pm=\mu_\sk \pm \frac{\Gamma_\sk}{2} \sqrt{\frac{k_\rA}{k_\sk|T|^2}-1} \, ,
   \label{convenient}
\ee
i.~e.~the skeleton chemical potential is the average of the two branches $\mu_\pm$ and the skeleton $\Gamma_\sk$ the (|T|-weighted) width
\be
    \Gamma_\sk = \left(\mu_+(|T|)-\mu_-(|T|)\right)\sqrt{\frac{|T|^2}{(k_\rA/k_\sk)-|T|^2}} \, .
\ee
Figure \ref{fig-comp03} shows a typical nonlinear Lorentz curve of the type (\ref{Lor_nl}) for the case of a repulsive nonlinearity. The skeleton curve $\mu_\sk$, indicated by the dashed line, appears as a kind of backbone structure of the nonlinear Lorentz profile, justifying its name which is taken from the theory of classical driven nonlinear oscillators (see e.g. \cite{Mick81}) where resonance curves similar to (\ref{Lor_nl}) occur.

The skeleton curves $\mu_\sk$ and $\Gamma_\sk$ can either be parametrized in terms of the amplitude $|C|^2$ or in terms of the number of particles $N=\int_{-b}^b |\psi_\sk(x)|^2 \rd x$ inside the potential well. It was shown (see e.g. \cite{Schl04,Schl06}) that in an adiabatic approximation the decay behavior of a resonance state is determined by the imaginary part $\Gamma_\sk  \left(N(t)\right)$ of the instantaneous chemical potential via
\be
   \partial_t N(t)=-\frac{\Gamma_\sk\left(N(t)\right)}{\hbar} N(t) \, .
   \label{N_dot0}
\ee
Thus there is a close connection between the transmission lineshape and the decay behavior of the corresponding resonance state as it is known for the linear limit $g=0$ where the decay coefficient is constant and the lineshape is Lorentzian.

\section{Calculating skeleton curves}
\label{sec_skel}
As shown in Sec.~\ref{sec-nl_Lorentz} the skeleton curves $\mu_\sk(|C|^2)$ and $\Gamma_\sk(|C|^2)$ are obtained by solving the NLSE (\ref{4}) with the Siegert boundary conditions (\ref{5a}), (\ref{5}). It has been shown that the use of Siegert boundary conditions is equivalent to a complex rotation of the coordinates (see e.g.~\cite{Mois98}). Different procedures based on this principle, e.g.~direct complex scaling or complex absorbing potentials, have been successfully applied to resonance states of the NLSE \cite{Schl04,Mois03,Wimb06}.
Here we present an alternative method which is numerically cheap, easy to implement and, though not quite as accurate as the complex scaling procedures, provides a convenient basis for approximations.

This method, which we call the Siegert method, is based upon neglecting the imaginary part $-\Gamma
_\sk/2$ of the chemical potential and thus having only real values of $k_\sk=\sqrt{2m(\mu_\sk-g|C|^2)}/\hbar$ which is justified for not too large values of $\Gamma_\sk$. Since the boundary conditions (\ref{5a}) and (\ref{5}) can no longer be satisfied simultaneously for real values of $k_\sk$, we replace the boundary conditions (\ref{5}) by the less restricting condition
\be
    \frac{\rd}{\rd x}|\psi_\sk|^2 \Big|_{x=0}=0
    \label{newBC}
\ee
which preserves the symmetry $|\psi_\sk(-x)|^2=|\psi_\sk(x)|^2$ of the skeleton wavefunctions.
The new boundary value problem given by (\ref{4}), (\ref{5a}) and (\ref{newBC}) is solved using a shooting procedure where the NLSE (\ref{4}) is integrated from $x=+a$ to $x=0$ using a Runge-Kutta solver with starting conditions (\ref{5a}) at $x=a$ for a fixed value of $C$. By means of a bisection method, $\mu_\sk$ is adapted until the condition (\ref{newBC}) is satisfied.
The imaginary part of the chemical potential can then be estimated by the Siegert relation
\be
   \Gamma_\sk=\frac{\hbar^2 k_\sk |C|^2}{\int_{0}^{b} |\psi_\sk(x)|^2 \rd x}
   \label{Gamma}
\ee
where $\pm b$ are the positions of the maxima of the symmetric trapping potential (see Appendix \ref{app_SiegertRel}).

Before we use our simple method to compute skeleton curves in Sec.~\ref{sec_app}, we demonstrate its validity for the lowest resonance state of the standard test potential
\be
   V(x)=\frac{x^2}{2}\exp(-\alpha x^2)
   \label{pot_0}
\ee
with $\alpha=0.1$ and $b\approx 3.16$ using units where $\hbar=1$ and $m=1$ as we do for all numerical calulations in this paper. We choose $a=30$ to be sufficiently large to ensure that the resonance wavefunction $\psi_\sk(x)$ is well approximated by a plane wave in the area $x>a$ . The amplitude $C$ is chosen such that the wavefunction is normalized in the region $|x|\le b$, i.e. $\int_{-b}^{b} |\psi_\sk(x)|^2 \rd x=1$. Note that because of the nonlinearity, $\psi_\sk$ is not proportional to $C$.

In Table \ref{tab-CS} we compare our results for the lowest resonance of the potential (\ref{pot_0}) with the results of direct
complex scaling and the complex absorbing potential method \cite{Schl04}. The agreement between the different methods is very good,
especially if the interaction constant $g$ is small which is also the case within our calculations of skeleton curves in the
following section. Apart from being numerically cheap and easy to implement the Siegert method proposed here can provide analytical expressions for $\mu_\sk$ and $\Gamma_\sk$ if the potential in consideration is simple enough (see subsection \ref{subsec_sq}).

\begin{table}[htbp]
\centering
\begin{tabular}{c|ccc|ccc}

 $g$ & $\mu_{\rm S}$ & $\mu_{\rm CS}$ & $\mu_{\rm CAP}$ &
  $\Gamma_{\rm S}/2$ & $\Gamma_{\rm CS}/2$& $\Gamma_{\rm CAP}/2$ \\
  \hline
  0 & 0.4601 & 0.4601 & 0.4602 & 9.63e-7 & 9.35e-7 & 9.62e-7 \\
  1 & 0.7954 & 0.7954 & 0.7954 & 1.81e-5 & 1.82e-5 & 1.80e-5 \\
  2 & 1.0772 & 1.0765 & 1.0772 & 1.56e-4 & 1.55e-4 & 1.56e-4 \\
  3 & 1.3192 & 1.3190 & 1.3192 & 8.11e-4 & 8.05e-4 & 8.05e-4 \\
  4 & 1.5317 & 1.5315 & 1.5312 & 2.79e-3 & 2.76e-3 & 2.75e-3 \\
  5 & 1.7247 & 1.7236 & 1.7231 & 6.78e-3 & 6.65e-3 & 6.63e-3 \\
  6 & 1.9070 & 1.9043 & 1.9035 & 1.27e-2 & 1.24e-2 & 1.23e-2
\end{tabular}
\caption{Chemical potential and decay rates for the lowest quasibound state, calculated with the Siegert method (S), complex scaling (CS) and complex absorbing potentials (CAP).}
\label{tab-CS}
\end{table}

\section{Applications}
\label{sec_app}
In this section we compare exact results for transmission peaks with the predictions of the nonlinear Lorentz profile (\ref{Lor_nl}) for two different model potentials. To this end we calculate the skeleton curves $\mu_\sk$ and $\Gamma_\sk$, which can either be parametrized in terms of the amplitude $|C|^2$ or in terms of the number of particles $N=\int_{-b}^b |\psi_\sk(x)|^2 \rd x$ inside the potential well, by means of the Siegert method presented in the previous section. Furthermore we show that these skeleton curves are conveniently approximated by polynomials depending on a few instructive parameters.

\subsection{Example 1: The double--Gaussian barrier}
\label{subsec_DGauss}

To demonstrate the validity of our model (\ref{Lor_nl}) we apply it to the potential
\be
   V(x)=V_0\left[\exp(-(x+b)^2/\alpha^2)+\exp(-(x-b)^2/\alpha^2)\right]
   \label{DGauss}
\ee
with the parameters $V_0=1$, $b=14.7/2$, $\alpha=b/5$ and a nonlinearity of $g=0.005$.
In \cite{Paul05} the transmission coefficient of this potential in dependence of $\mu$ is calculated  for the case of an initially empty waveguide. The incoming amplitude $|A|^2$ is connected with the incident current $j_{\rm in}$ (i.e.~the current in absence of the barrier) via $j_{\rm in} =|A|^2\sqrt{2(\mu-g|A|^2)/m}$. In the following we will assume $|A|^2=1$ in all numerical calculations.

Using the method described in Sec.~\ref{sec_skel} we numerically calculate the skeleton curves $\mu_\sk(|C|^2)$, $\Gamma_\sk(|C|^2)$ with $|C|^2 \le |A|^2$. For $|C|^2=|A|^2$ we have $|T|^2=1$ (see Sec.~\ref{sec-nl_Lorentz}). We call the quantities $\mu_R:=\mu_\sk(|C|^2=|A|^2)$,
$\Gamma_R:=\Gamma_\sk(|C|^2=|A|^2)$ and $\psi_R(x):=\psi_\sk(x; |C|^2=|A|^2)$ the resonance chemical potential, resonance width and resonance wavefunction respectively.
In the limit $|C|^2 \rightarrow 0$ the influence of the nonlinear term in the NLSE (\ref{4}) can be neglected so that $\mu_\sk(|C|^2 \rightarrow 0) =\mu_n$ and $\Gamma_\sk(|C|^2 \rightarrow 0)=\Gamma_n$ where $\mu_n$ and $\Gamma_n$ are the respective quantities of the linear problem with $g=0$.

\begin{figure}[htb]
\centering
\includegraphics[width=8cm, angle=0]{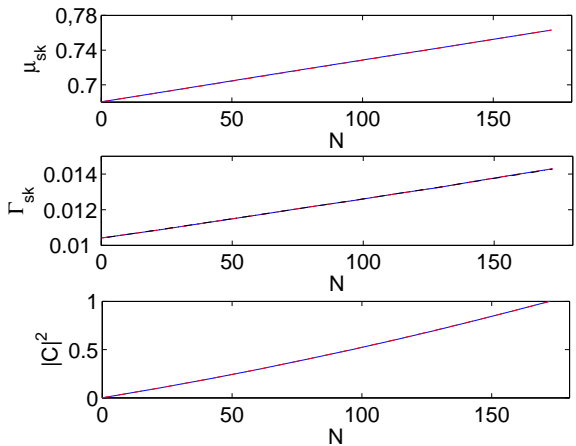}
\caption{\label{fig-comp02} {(Color online) The numerically calculated curves $\mu_\sk(N)$, $\Gamma_\sk(N)$ and $|C|^2(N)$ (solid blue) and the approximation described by (\ref{musk_N}) -- (\ref{NR}) (dashed red) for the potential (\ref{DGauss}) as well as the Taylor approximation (\ref{Gamma_Taylor}) (dashed dotted black curve) are almost indistinguishable on the scale of drawing.
}}
\end{figure}
\begin{figure}[htb]
\centering
\includegraphics[width=7cm,  angle=0]{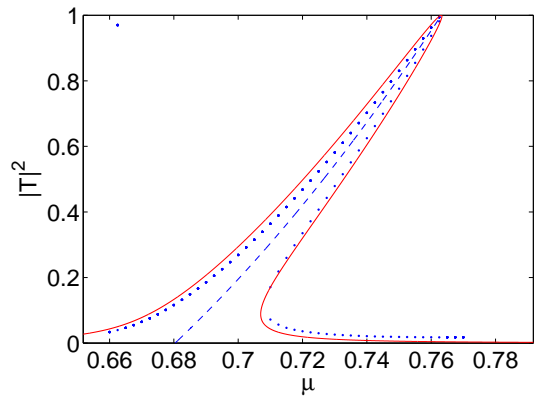}
\caption{\label{fig-comp03} {(Color online) Comparison between the nonlinear Lorentz curve (\ref{Lor_nl}) (solid red curve) and the transmission coefficient obtained from solving the stationary NLSE (dotted blue curve) for the double Gaussian potential (\ref{DGauss}). Dashed blue: Skeleton curve $\mu_\sk(|T|^2)$.
}}
\end{figure}

Now we will show that over a wide range of parameters the skeleton curves $\mu_\sk(N)$ and $\Gamma_\sk(N)$ are well approximated by simple elementary functions and that the five quantities $\mu_n$, $\Gamma_n$, $\mu_R$, $\Gamma_R$ and $|A|^2$ provide all the necessary information.

Since the shift in the chemical potential is caused by the term $g|\psi_\sk|^2$ in the NLSE (\ref{4}), we assume this shift to be approximately proportional to the number of particles inside the potential well, that is
\be
    \mu_\sk(N)=\mu_n+\frac{\mu_R-\mu_n}{N_R}N
    \label{musk_N}
\ee
where $N_R=\int_{-b}^{b}|\psi_R(x)|^2 \rd x$ is the norm inside the well in the case of resonance.
Next we represent the amplitude $|C|^2=|\psi(a)|^2$ as a function of $N$ by a Taylor series which we truncate after the quadratic term,
\be
   |C|^2 \approx |T|^2|A|^2 \approx \chi_1 N+\chi_2 N^2 \, .
   \label{Cq_N}
\ee
If there are no particles ($N=0$), the transmitted amplitude $|C|^2$ is zero so that there is no constant term in (\ref{Cq_N}).
Inserting (\ref{musk_N}) and (\ref{Cq_N}) into (\ref{Gamma}) we obtain
\begin{eqnarray}
   \Gamma_\sk(N)=\frac{2 \hbar^2 k_\sk|C|^2}{N} \quad\quad\quad\quad\quad\quad\quad\quad\quad\quad\quad\quad\quad \label{Gamma_N}\\ \nonumber
    \quad \approx\frac{2\sqrt{2}\hbar}{\sqrt{m}}\sqrt{\mu_\sk(N)-g(\chi_1 N+\chi_2 N^2)}(\chi_1 +\chi_2 N).
\end{eqnarray}
From Eq.~(\ref{Gamma_N}) we obtain
\begin{eqnarray}
\Gamma_\sk(N=0)&=&\Gamma_n=2\sqrt{2}\hbar\sqrt{\mu_n/m}\chi_1 \\ \label{NR}
N_R&=&2 \hbar \sqrt{2}\sqrt{\mu_R-g|A|^2}|A|^2/(\sqrt{m}\Gamma_R) \quad
\end{eqnarray}
so that the coefficients $\chi_1$ and $\chi_2$ are given by
$\chi_1=\Gamma_n\sqrt{m}/(2 \hbar \sqrt{2 \mu_n}) $ and $\chi_2=|A|^2/N_R^2-\chi_1/N_R$.
Inverting Eq.~(\ref{Cq_N}) leads to
$N_\pm=-\chi_1/(2 \chi_2)\pm\sqrt{\chi_1^2/(4 \chi_2^2)+|C|^2/\chi_2}$. Thus we can compute
$\mu_\sk$ and $\Gamma_\sk$ as a function of $|T|^2$.

It is often useful to approximate Eq.~(\ref{Gamma_N}) by a second order Taylor polynomial
\be
   \Gamma_\sk=\Gamma_n \left(1+ \eta_1 N +\eta_2 N^2  \right)
   \label{Gamma_Taylor}
\ee
where $\eta_1=\chi_2/\chi_1+(\mu_R-\mu_n)/(2 \mu_n N_R)-g\chi_1/\mu_n$ and $\eta_2=\chi_2(\mu_R-\mu_n)/(2 \mu_n N_R \chi_1)-g \chi_1/\mu_n-\left[(\mu_R-\mu_n)/(\mu_n N_R)-g\chi_1/\mu_n \right]^2$.

Figure \ref{fig-comp02} reveals an excellent agreement between the numerically calculated skeleton curves and the approximation described by (\ref{musk_N}) -- (\ref{NR}).

Figure \ref{fig-comp03} compares a peak of the transmission coefficient with the resonance model (\ref{Lor_nl}). While the qualitative features such as the bending of the curve are well reproduced, the resonance model (\ref{Lor_nl}) slightly overestimates the width of the resonance curve.

\subsection{Example 2: The square well}
\label{subsec_sq}
For illustrative purposes we now apply the result (\ref{Lor_nl}) to a simple analytically solvable toy model system which has a similar transmission behavior as the double--Gaussian barrier considered in the previous section. In addition it shows resonance peaks originating from the bound states of the corresponding linear ($g=0$) system which have been destabilized due to repulsive ($g>0$) interaction and have thus undergone a transition from bound to resonance state \cite{06nl_transport}.
We consider the finite square well potential with vanishing interaction outside the potential well,
where the wavefunction $\psi(x)$ must satisfy
\be
   \left( -\frac{\hbar^2}{2m} \frac{d^2}{dx^2} -\mu \right) \psi(x) = 0 , \quad |x|>a
   \label{sq_aussen}
\ee
and
\be
    \left( -\frac{\hbar^2}{2m} \frac{d^2}{dx^2} + g|\psi(x)|^2 +V_0 -\mu \right) \psi(x) =
    0,  \quad  |x|\le a \, .
    \label{sq_innen}
\ee
This model with vanishing interaction outside the potential well was introduced in \cite{06nl_transport} in order to discuss the scattering process in terms of ingoing and outgoing waves and thus enabling an analytical treatment.
In principle the interaction outside the potential well can be eliminated by means of a magnetic Feshbach resonance (see e.~g.~\cite{Volz03}) or by a larger transversal extension $a_\perp$ of the waveguide in this region since the effective one--dimensional interaction strength is proportional to $1/a_\perp^2$ (see e.~g.~\cite{Pita03}). Alternatively, instead of neglecting the interaction outside the potential well one might add additional repulsive barriers at $x=\pm a$ without affecting much the qualitative behavior of the system with the disadvantage of making the analytical treatment more complicated.

In \cite{06nl_transport} the transmission coefficient is calculated analytically and it is shown that the respective states satisfying the NLSE (\ref{sq_innen}), (\ref{sq_aussen}) with the boundary conditions (\ref{5a}) and (\ref{newBC}) (skeleton states) have the chemical potential
\be
   \mu_\sk=V_0+\frac{3}{2}g|C|^2+\frac{\hbar^2 K^2(p) n^2}{2 m a^2}\left\{ \begin{array}{cl} 1+p\\1-2p \end{array} \right.
   \label{sq_mu}
\ee
where $n$ is an integer number and $K(p)$ is the complete elliptic integral of the first kind (see e.g.~\cite{Abra72}). The upper and lower alternative correspond to $g|C|^2(V_0+g|C|^2) \ge 0$ and $0\le g|C|^2\le |V_0|$ respectively.
The parameter $0 \le p \le 1$ is determined by
\be
  g|A|^2[V_0+g|C|^2]\frac{2 m^2 a^4}{\hbar^4 n^4}=K^4(p)\left\{ \begin{array}{cl} p\\p(p-1) \end{array} \right.
\ee
and the norm of the wavefunction inside the well reads
\be
   N=2a|C|^2+\frac{2 \hbar^2 K(p) n^2}{g m a}\left\{ \begin{array}{cl} K(p)-E(p)\\(1-p)K(p)-E(p) \end{array} \right. \, .
   \label{sq_N}
\ee
Since $g=0$ in the region $|x|>a$ we have $k_\sk=\sqrt{2m \mu_\sk}/\hbar$ so that the decay width is given by
\be
    \Gamma_\sk=\frac{2 \hbar \sqrt{2 \mu_R} |C|^2}{\sqrt{m} N} \, ,
\ee
where $\mu_\sk$ and $N$ are given in (\ref{sq_mu}) and (\ref{sq_N}).
As in Sec.~\ref{subsec_DGauss}, the skeleton curves can be approximated by a Taylor polynomial.
Applying our model to resonances of the square well potential (\ref{sq_aussen}), (\ref{sq_innen})
it turns out that it is sufficient to truncate the Taylor polynomials in (\ref{Cq_N}) and (\ref{Gamma_Taylor}) after the linear term. This leads to $N/N_R=|T|^2$ and thus
\begin{eqnarray}
   \mu_\sk&=&\mu_n+(\mu_R-\mu_n)|T|^2 \,, \label{mu_Gamma_1}\\
    \Gamma_\sk&=&\Gamma_n+(\Gamma_R-\Gamma_n)|T|^2 \, .
   \label{mu_Gamma_2}
\end{eqnarray}


\begin{figure}[htb]
\centering
\includegraphics[width=7cm,  angle=0]{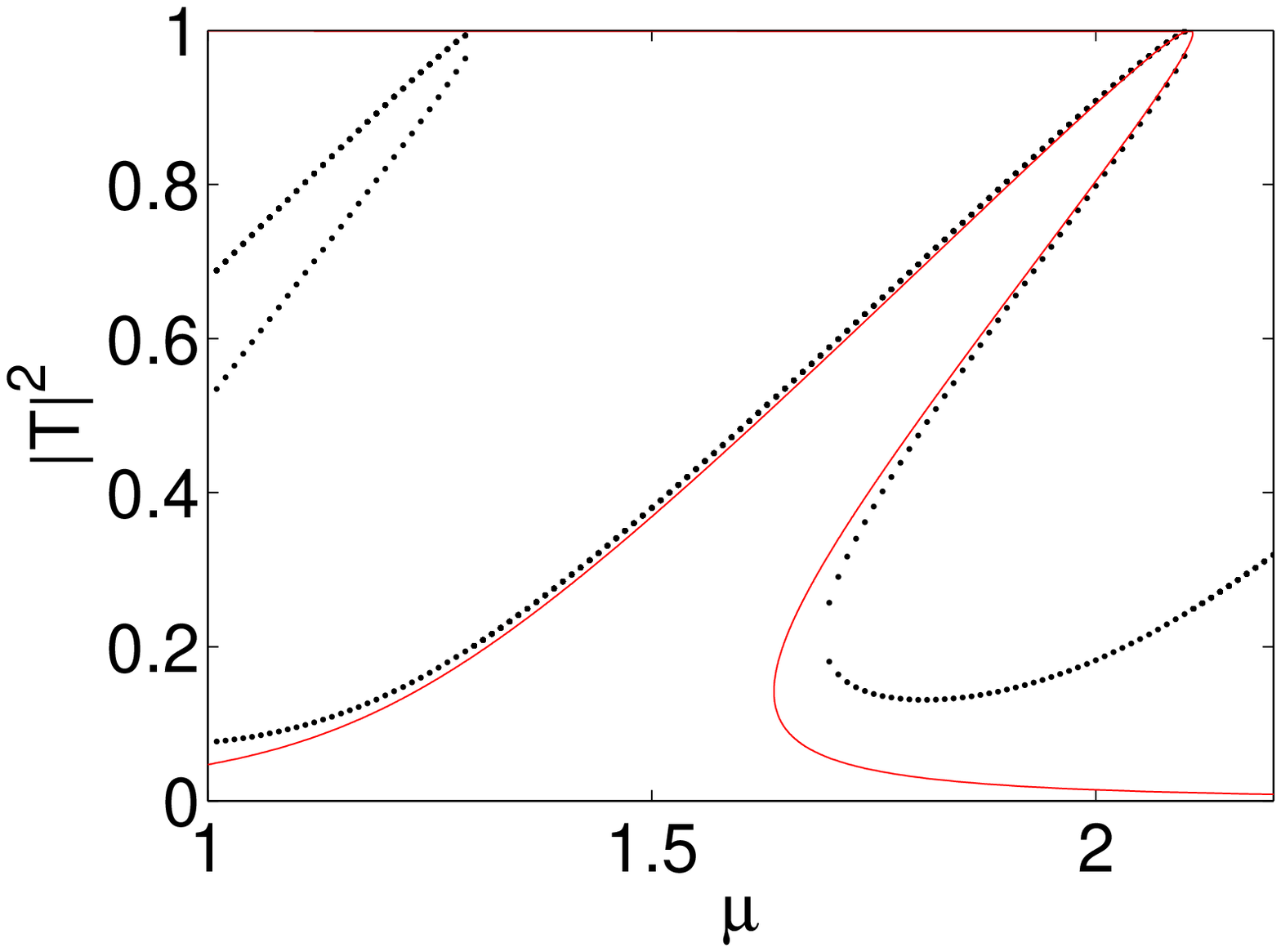}
\includegraphics[width=7cm,  angle=0]{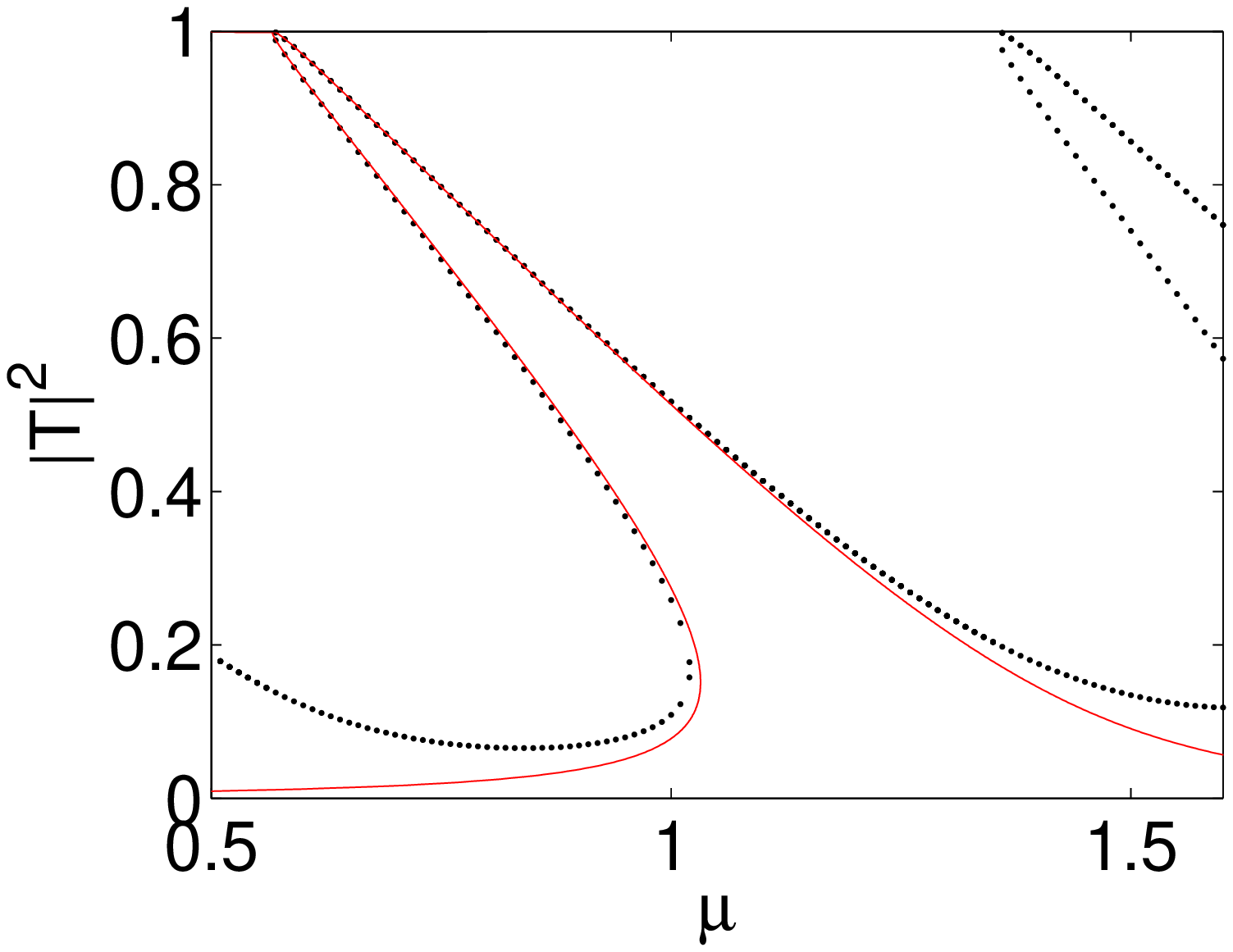}
\caption{\label{fig-comp} {(Color online) Nonlinear Lorentz curve (\ref{Lor_nl}) (solid red curve) and  transmission coefficient obtained from solving the stationary NLSE (dotted black curve) for a square well potential with  $a=20$, $V_0=-50$ for the resonance with quantum number $n=129$ for $g=+1$ (upper panel) and $g=-1$ (lower panel).
}}
\end{figure}
\begin{figure}[htb]
\centering
\includegraphics[width=7cm,  angle=0]{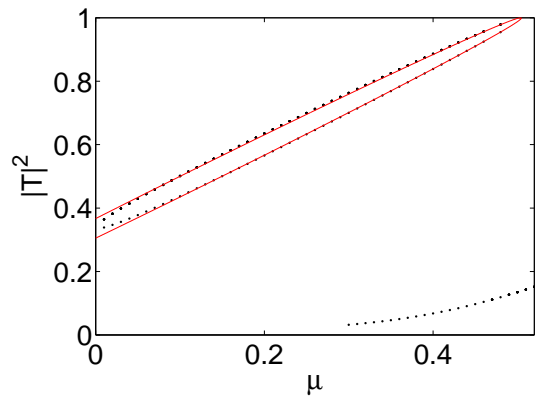}
\caption{\label{fig-comp2} {(Color online) Nonlinear Lorentz curve (\ref{Lor_nl}) (solid red curve) and  transmission coefficient obtained from solving the stationary NLSE (dotted black curve) for a square well potential with  $a=20$, $V_0=-50$ for the resonance with quantum number $n=127$, which had been a bound state in the linear case $g=0$,  for $g=+1$.
}}
\end{figure}

Figures \ref{fig-comp} and \ref{fig-comp2} show the transmission probability $|T|^2(\mu)$ in the vicinity of a resonance, the exact solution and the resonance approximation introduced in Sec.~\ref{sec_skel} for a deep square well with $V_0=-50$ and $a=20$.
For both repulsive and attractive interaction, where the curves bend to the right or left, respectively, a good agreement between the nonlinear Lorentz curve (\ref{Lor_nl}) in first order approximation (\ref{mu_Gamma_1}), (\ref{mu_Gamma_2}) and the respective resonance peak (see \cite{06nl_transport}) is observed.
In particular, Fig. \ref{fig-comp2} shows that the nonlinear Lorentz curve (\ref{Lor_nl}) is also able to describe the unusually shaped peaks surrounding resonances which correspond to bound states in the linear limit $g \rightarrow 0$ (see \cite{06nl_transport}).
The deviations are due to the fact that only a single resonance is included in the present approximation.

\subsection{Decay behavior}

As discussed in section \ref{sec-nl_Lorentz} the decay behavior of the resonance state is described by
\be
   \partial_t N=-\frac{\Gamma_\sk(N)}{\hbar} N \, .
   \label{N_dot}
\ee
For the simple situation when the skeleton curves are approximately linear in $|T|^2$, as in (\ref{mu_Gamma_1}), (\ref{mu_Gamma_2}), an analytical expression for the time dependence can be derived.
With $\Gamma_\sk(N)=\Gamma_n +(\Gamma_R-\Gamma_n)(N/N_R)$ we obtain
\be
   \hbar \partial_t \frac{N}{N_R}=-\Gamma_n \frac{N}{N_R}-(\Gamma_R-\Gamma_n)\left( \frac{N}{N_R} \right)^2 \, .
\ee
Separation of variables yields
\be
     - \frac{\hbar \rd y}{\Gamma_n y+ (\Gamma_R-\Gamma_n)y^2}=\rd t
\ee
with $y=N/N_R$. This can be integrated to give
\be
   N(t)=\frac{\Gamma_n N_R}{\Gamma_R (\re^{\Gamma_n(t-t_0)/\hbar}-1)+\Gamma_n } \, .
   \label{decay_Lorentz2}
\ee

In the limit $g \rightarrow 0$ this reduces to the linear decay behavior $N(t)=N_R \exp[-\Gamma_n (t-t_0)/\hbar]$. In the limit of long times $t\gg t_0$ the system shows a linear decay $N(t) \rightarrow [\Gamma_n/\Gamma_R]N_R\exp[-\Gamma_n (t-t_0)/\hbar] \sim \exp[-\Gamma_n (t-t_0)/\hbar]$ as well.

\begin{figure}[htb]
\centering
\includegraphics[width=7cm,  angle=0]{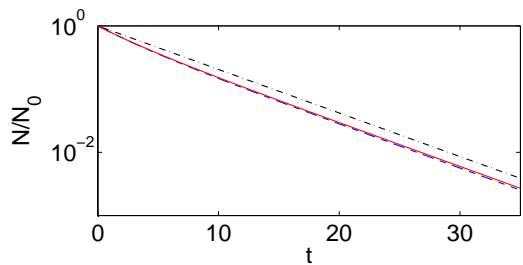}
\caption{\label{fig-decay} {(Color online) Decay according to formula (\ref{decay_Lorentz2}) (solid red line), the numerical solution of Eq.~(\ref{N_dot}) (dashed blue line) and the linear decay behavior (dashed-dotted black line) for the parameters $a=20$, $V_0=-50$ and $g=+2$ .
}}
\end{figure}

Figure \ref{fig-decay} shows the decay according to formula (\ref{decay_Lorentz2}) and compares it with the numerical solution of Eq.~(\ref{N_dot}) and the linear decay in a semilogarithmic plot. Formula  (\ref{decay_Lorentz2}) agrees well with the numerical solution of Eq.~(\ref{N_dot}).
In the limit of long times both curves are parallel to the linear decay curve so that the system adopts a linear decay behavior as predicted above.

%

\section{Conclusion}
We have presented an analysis of the nonlinear resonances found
for transmission of a BEC through a one-dimensional potential
barrier in a mean-field GPE description. The Siegert method for
determination of resonances is generalized to the nonlinear case
providing a convenient recipe for the computation of nonlinear
resonances.

Based on this Siegert method, we developed a formula for the
nonlinear Lorentz profiles which can be described in terms of skeleton
functions depending on a few instructive parameters. The skeleton curves also determine the decay behavior of the underlying resonance state thus relating the transmission lineshape to the resonance lifetime.

Applications to a double Gaussian barrier and a square well potential illustrate
and support our analysis. Finally, for a simple model an analytical expression for the decay behavior could be derived. We are therefore hopeful that the
theoretical ideas presented may be useful in future work
on nonlinear resonances.

\begin{acknowledgments}
Support from the Deutsche
Forschungsgemeinschaft via the Graduiertenkolleg ''Nichtlineare Optik
und Ultrakurzzeitphysik'' is gratefully acknowledged. We also
thank Tobias Paul, Peter Schlagheck and Dirk Witthaut for valuable
discussions.
\end{acknowledgments}

\begin{appendix}
\section{Analytical continuation}
\label{app_Continuation}

Since the NLSE
\be
-\frac{\hbar^2}{2m}\psi''(x)+V(x)\psi(x)+g|\psi(x)|^2\psi(x)=\mu \psi(x)
\label{GPE_AC}
\ee
explicitly contains the squared magnitude $|\psi(x)|^2$ of the wavefunction it is not analytical and therefore the analytical continuation of its solutions for complex values of $\mu$ is nontrivial.
Following the arguments given in \cite{Cart08} we decompose the solution of the stationary GPE as $\psi(x)=X(x)+\ri Y(x)$ with real functions $X(x)$ and $Y(x)$.
From the real and imaginary part of Eq.~(\ref{GPE_AC}) we get a system of two equations
\be
    -\frac{\hbar^2}{2m}X''(x)+V(x)X(x)+g(X^2(x)+Y^2(x))X(x)=\mu X(x) \label{X}
\ee
\be
    -\frac{\hbar^2}{2m}Y''(x)+V(x)Y(x)+g(X^2(x)+Y^2(x))Y(x)=\mu Y(x) \label{Y}
\ee
which are analytical. The solutions of this system of equations therefore have a straightforward continuation into the domain of complex chemical potentials.
As an example, we consider the plane wave solution of the free ($V(x)=0$) GPE with $\psi(x)=C \exp(\ri k_C x)$, $k_C=\sqrt{2m(\mu-g|C|^2)}/\hbar$ and $C=|C|\exp(\ri \theta)$. One can easily verify that its decomposition $X(x)=|C|\cos(k_Cx+\theta)$, $Y(x)=|C|\sin(k_Cx+\theta)$ satisfies the system (\ref{X}), (\ref{Y}) for all complex values of $\mu$.

\section{The Siegert relation}
\label{app_SiegertRel}
\noindent
{\bf Derivation:}
In the following we will derive a formula for the decay coefficient of a resonance state of an arbitrary symmetric finite range potential. For the sake of generality and for future applications we consider the cases of one, two and three dimensions simultaneously. For now we only consider resonances of the linear Schr\"odinger equation (Eq.~(\ref{GPE}) with $g=0$). The applicability to the nonlinear case is discussed separately further below.

Any solution $\psi({\bf x},t)=\sqrt{\rho({\bf x},t)}\exp[\ri \phi({\bf x},t)]$ of the Schr\"odinger equation (\ref{GPE}) with real functions $\phi({\bf x},t)$ and $\rho({\bf x},t)>0$ satisfies the continuity equation
\be
   \partial_t \rho+ {\rm div} {\bf j} =0
\ee
where
\be
  {\bf j}=\frac{\hbar}{m} \rho \nabla \phi \, .
\ee
Application of the Gauss theorem for vector fields leads to
\be
    \partial_t N=-\int_{A} {\bf j} \cdot \rd {\bf A}
    \label{NPunkt}
\ee
where
\be
  N=\int_{\cal{V}} \rho({\bf x},t) \, \rd^D {\bf x}
  \label{N}
\ee
is the norm of the wavefunction within an $D$-dimensional volume $\cal{V}$ and ${\bf A}=\partial(\cal{V})$ is the directed surface of $\cal{V}$.
If the wavefunction is trapped inside the volume $\cal{V}$ the decay coefficient can be defined by the relation
\be
   \partial_t N=-\frac{\Gamma}{\hbar} N \, .
\ee
Together with Eqs.~(\ref{NPunkt}) and (\ref{N}) this leads to
\be
    \Gamma=-\hbar\frac{\partial_t N}{N}=\hbar\frac{\int {\bf j} \cdot \rd {\bf A}}{\int_{\cal{V}} \rho({\bf x},t) \, \rd^D {\bf x}} \, .
    \label{Gamma_3D}
\ee

Now we consider a radially symmetric potential
$V({\bf x})=V(r)$ with finite range $a$, i.e. $V(r)=0$ if $r>a$, where $r=|\bf{x}|$. We assume the potential $V(r)$ to have a single maximum located at $b \le a$. Assuming that the wavefunction varies slowly in time, we replace the time--dependent wavefunction $\psi(\bf x,t)$ by the adiabatic resonance state $\psi(\bf x)$ of the stationary Schr\"odinger equation ( Eq.~(\ref{NLSE_stat}) with $g=0$).
Due to symmetry, the wavefunction in the area $r \le a$ can be written in polar coordinates as
\be
   \psi({\bf x})=R(r)Y(\Omega) \exp \left[\ri \phi_r(r)+\ri \phi_{\Omega}(\Omega)\right], \, r\le a,
   \label{WF_innen}
\ee
with real functions $R(r)$ and $Y(\Omega)$ where $\Omega$ stands for the angle variables.
The resonance wavefunctions of such a potential are obtained by applying purely outgoing (Siegert) boundary conditions,
\begin{eqnarray}
  \psi({\bf x}) &=& R(a) \left(\frac{a}{r} \right)^{(D-1)/2} Y(\Omega) \label{ausl_Welle} \\ \nonumber
  &\times&  \exp \left[{\ri k (r-a) +\ri \phi_r(a)+\ri \phi_\Omega(\Omega)}\right], \, r \ge a,
\end{eqnarray}
where $k={\rm Re}(\sqrt{2m \mu}/\hbar)$. For narrow resonances where $\Gamma/2=-{\rm Im}(\mu)$ is small compared to $\epsilon={\rm Re}(\mu)$ we can make the approximation $k \approx \sqrt{2m {\rm Re}(\mu)}/\hbar$. For $D=1,3$ (\ref{ausl_Welle}) is an exact solution, for $D=2$ it only holds in the limit $a \rightarrow \infty$ (see below). This ansatz makes the wavefunction continuous at $r=a$. The continuity of the derivative implies the conditions
\be
   R'(a)=\frac{1-D}{2a}R(a), \quad \quad \phi_r'(a)=k,
   \label{RB}
\ee
where the prime denotes the partial derivative with respect to $r$.

The resonance wavefunction shall be trapped in the region $0 \le r \le b$.
Thus the volume $\cal{V}$ is a $D$-dimensional sphere with radius $b$ so that $\rd^D{\bf x}=r^{D-1}\rd r \rd \Omega$ and
$\rd {\bf A}= {\bf e_r} b^{D-1} \rd \Omega$ where ${\bf e_r}$ is a unit vector in the radial direction.
For the integral (\ref{NPunkt}) we need the scalar product ${\bf e_r} \cdot {\bf j} =(\hbar/m) R^2(r)Y^2(\Omega){\bf e_r} \cdot \nabla (\phi_r+\phi_\Omega)$.
The volume integral (\ref{N}) becomes
\be
   N=\int_{0}^b R^2(r) r^{D-1} \rd r \int_{\Omega} Y^2(\Omega) \rd \Omega \, .
   \label{N1D}
\ee
For the surface integral (\ref{NPunkt}) we make the approximation
\be
   -\partial_t N=\int_{A} {\bf j} \cdot \rd {\bf A} \approx \int_{A'} {\bf j} \cdot \rd {\bf A'}
\ee
where ${\bf A'}$ is the surface of the sphere with radius $a$ which means that the reflection in the region $b \le r \le a$ is neglected and (\ref{NPunkt}) becomes
\be
   -\partial_t N\approx\int_{A'} {\bf j} \cdot \rd {\bf A'} = \frac{\hbar}{m} a^{D-1} R^2(a) \phi_r'(a) \int_{\Omega} Y^2(\Omega) \rd \Omega \, .
   \label{NPunkt1D}
\ee
By inserting Eqs.~(\ref{NPunkt1D}), (\ref{N1D}) and (\ref{RB}) into (\ref{Gamma_3D}) we finally obtain the formula
\be
   \Gamma = \frac{\hbar^2 k}{m} \frac{R^2(a) a^{D-1}}{\int_{0}^b R^2(r) r^{D-1} \rd r}
   \label{Gamma_b}
\ee
for the decay coefficient of a resonance with the chemical potential $\mu$ of the potential $V(r)$ with the finite range $a$ which is trapped inside the region $0\le r \le b$.

The formula (\ref{Gamma_b}) for $D=1$ resembles Eq.~(\ref{15}). In contrast to formula (\ref{Gamma_b}), the wavenumber $k_\sk$ in Eq.~(\ref{15}) is complex and the integration extends over the whole region $r \le a$. Thus Eq.~(\ref{Gamma_b}) can be regarded as an approximation to Eq.~(\ref{15}) (respectively to its generalization to higher dimensions (see \cite{Sieg39})). \\ [1mm]

\noindent
In the {\bf two--dimensional case}
the ansatz (\ref{ausl_Welle}) is only an approximation. As promised above we discuss this in more detail now by inserting the ansatz
\be
   \psi(r)=R(r) \exp(\ri \phi_r(r))
   \label{Ansatz_2D}
\ee
with real functions $R(r)$ and $\phi(r)$ into the radial part
\be
   \frac{\partial^2 \psi(r)}{\partial r^2}+\frac{1}{r} \frac{\partial \psi(r)}{\partial r} +\frac{2 m \mu}{\hbar^2}\psi(r)=0
   \label{SE_2D}
\ee
of the two--dimensional Schr\"odinger equation. Separating real and imaginary parts we arrive at
\begin{eqnarray}
   \frac{R''(r)}{R(r)}+\frac{R'(r)}{r R(r)}-\phi_r'^2(r)+\frac{2m \mu}{\hbar^2} &=&0 , \\
   \left(\frac{2R'(r)}{R(r)}+\frac{1}{r}\right)\phi_r'(r)+\phi_r''(r)&=&0 \,. \\ \nonumber
\end{eqnarray}
The choice $R(r)=C/\sqrt{r}$ and $\phi_r(r)=k r +\phi_0$ with real constants $C$, $k$ and $\phi_0$ solves the lower equation. The remaining equation yields $k^2=2m \mu/\hbar^2+r^{-2}/4$. Thus Eq.~(\ref{SE_2D}) is approximately solved by $\psi(r)=C/\sqrt{r} \exp(\ri k r +\phi_0)$ with $k=\sqrt{2 m \mu}/\hbar$ if $r^2 \gg \hbar^2/(8m \mu)$ so that the length $a$ in Eq.~(\ref{ausl_Welle}) must be chosen accordingly. \\

\noindent
{\bf Applicability to the NLSE:}
In the case of the NLSE (\ref{NLSE_stat}), the ansatz (\ref{WF_innen}), (\ref{ausl_Welle}) and thus formula (\ref{Gamma_b}) are still valid in many cases where the wave equation can still be separated into a radial part and an angular part.

For $D=1$, inserting the ansatz (\ref{WF_innen}), (\ref{ausl_Welle}) into the nonlinear term in (\ref{NLSE_stat}) leads to $g|\psi({\bf x})|^2=g R^2(r) Y^2(\Omega)=g R^2(r)$ since $Y(\Omega)=\pm 1$. This means that the nonlinear term only modifies the radial part of the wave equation whereas the angular part is not affected.  For $r \ge a$, $g|\psi({\bf x})|^2=g R^2(a)$ is a constant term which only causes a shift in the chemical potential. If $g R^2(a)< {\rm Re}(\mu)$ this can be accounted for by replacing $\mu \rightarrow \mu-gR^2(a)$ so that the wavevector is now given by $k={\rm Re}(\sqrt{2m(\mu-gR^2(a))}/\hbar)$.

If $D=2$, we also have $|Y(\Omega)|= 1$, so that the wavefunction can still be separated into a radial part and an angular part in analogy to the one--dimensional case. For $r \ge a$, $g|\psi({\bf x})|^2=g R^2(a)a/r$. Thus for $|g| R^2(a) \ll |\mu|$ we can neglect the influence of the nonlinear term in the region $r \ge a$ and the ansatz (\ref{ausl_Welle}) is still valid.

For $D=3$, we have $|Y(\Omega)| \ne 1$ in general so that the wave equation can no longer be separated into a radial part and an angular part an thus the ansatz (\ref{WF_innen}), (\ref{ausl_Welle}) fails. However, for the special case of $s$--wave solutions $|Y(\Omega)|= 1$ still holds and in analogy to the case $D=2$, formula (\ref{Gamma_b}) is approximately valid if $|g| R^2(a) \ll |\mu|$.


\end{appendix}

\end{document}